\documentstyle[prl,aps,multicol]{revtex}
\renewcommand{\narrowtext}{\begin{multicols}{2} \global\columnwidth20.5pc}
\renewcommand{\widetext}{\end{multicols} \global\columnwidth42.5pc}
\begin{document}
\draft
\title{{\bf Theory\ of\ random\ matrices with strong level confinement:}\\
{\bf orthogonal polynomial approach}}
\author{{\bf V. Freilikher}$\dagger $, {\bf E. Kanzieper}$\dagger $,{\bf \ and I.
Yurkevich}$\ddagger ^{*}$}
\address{$\dagger $The Jack and Pearl Resnick Institute of Advanced Technology,\\
Department of Physics, Bar-Ilan University, Ramat-Gan 52900, Israel\\
$\ddagger $International Centre for Theoretical Physics, Trieste 34100, Italy}
\date{April 14, 1996}
\maketitle

\begin{abstract}
Strongly non-Gaussian ensembles of large random matrices possessing unitary
symmetry and logarithmic level repulsion are studied both in presence and
absence of hard edge in their energy spectra. Employing a theory of
polynomials orthogonal with respect to exponential weights we calculate with
asymptotic accuracy the two-point kernel over all distance scale, and show
that in the limit of large dimensions of random matrices the properly
rescaled local eigenvalue correlations are independent of level confinement
while global smoothed connected correlations depend on confinement potential
only through the endpoints of spectrum. We also obtain exact expressions for
density of levels, one- and two-point Green's functions, and prove that new
universal local relationship exists for suitably normalized and rescaled
connected two-point Green's function. Connection between structure of
Szeg\"o function entering strong polynomial asymptotics and mean-field
equation is traced.
\end{abstract}

\pacs{\tt cond-mat/9604078} 

\narrowtext

\section{Introduction}

Statistical properties of complex physical systems can successfully be
investigated within the framework of random-matrix theory (RMT) \cite{Mehta}%
. It turned out to be quite general and powerful phenomenological approach
to description of various phenomena in such diverse fields as
two-dimensional gravity \cite{Gravity}, quantum chaos \cite{Chaos}, complex
nuclei \cite{Nuclei} and mesoscopic physics \cite{Mesoscopics}.

In all the realms mentioned above the physical systems can be described with
the help of different matrix models whose structures depend on physical
properties of the systems involved. In the applications of RMT to the
complex quantum-mechanical objects the real Hamiltonian is rather intricate
to be handled or simply unknown. In such situation the integration of exact
equations is replaced by the study of the joint distribution function $%
P\left[ {\bf H}\right] $ of the matrix elements of Hamiltonian ${\bf H}$. If
there is not preferential basis in the space of matrix elements (i.e. the
system in question is ``as random as possible'', and equal weight is given
to all kinds of interactions) one has to require $P\left[ {\bf H}\right]
d\left[ {\bf H}\right] $ to be invariant under similarity transformation $%
{\bf H}\rightarrow {\cal R}^{-1}{\bf H}{\cal R}$ with ${\cal R}$ being
orthogonal, unitary or symplectic $n\times n$ matrix reflecting the
fundamental symmetry of the underlying Hamiltonian. The general form of $%
P\left[ {\bf H}\right] $ compatible with invariance requirement is 
\begin{equation}
P\left[ {\bf H}\right] =Z^{-1}\exp \left\{ -\text{tr}V\left[ {\bf H}\right]
\right\}  \label{i.00}
\end{equation}
with arbitrary $V\left[ {\bf H}\right] $ providing existence of partition
function $Z$. Introducing matrix ${\cal S}_\beta $ that diagonalizes
Hamiltonian ${\bf H}$, ${\bf H}={\cal S}_\beta ^{-1}{\bf X}{\cal S}_\beta $,
and carrying out the integration over orthogonal $\left( \beta =1\right) $,
unitary $\left( \beta =2\right) $ or symplectic $\left( \beta =4\right) $
group $d\mu \left( {\cal S}_\beta \right) $ in the construction $P\left[ 
{\bf H}\right] d\left[ {\bf H}\right] $, one obtains the famous expression
for the joint probability density function of the eigenvalues $\left\{
x\right\} $ of the matrix ${\bf H}$: 
\begin{equation}
P\left( \left\{ x\right\} \right) =Z^{-1}\exp \left\{ -\beta \left[
\sum_iV\left( x_i\right) -\sum_{i<j}\ln \left| x_i-x_j\right| \right]
\right\} \text{.}  \label{i.01}
\end{equation}
Level repulsion described by the logarithmic term is originated from
Jacobian $\prod_{i<j}\left| x_i-x_j\right| ^\beta $ arising when passing
from the integration over independent elements $H_{ij}$ of the Hamiltonian $%
{\bf H}$ to the integration over smaller space of its $n$ eigenvalues $%
\left\{ x\right\} $. Confinement potential $V\left( x\right) $, which
determines (together with logarithmic law of level repulsion) the mean level
density, contains an information about correlations between different matrix
elements of a random Hamiltonian ${\bf H}$. [Note, that parameter $\beta $
is factored out from $V\left[ {\bf H}\right] $ in Eq. (\ref{i.00}) to fix
density of levels in random-matrix ensembles with the same confinement
potential but with different underlying symmetries].

In the matrix formulation given above the eigenvalues $\left\{ x\right\} $
of the Hamiltonian ${\bf H}$ run from $-\infty $ to $+\infty $. Formally,
the same matrix model Eq. (\ref{i.01}) appears in so-called maximum entropy
models constructed to describe transport properties of mesoscopic systems.
In this case there is additional positivity constraint on $\left\{ x\right\} 
$, $x\geq 0$, that directly follows from unitarity of scattering matrix \cite
{Mesoscopics,Slevin-Nagao} and introduces the hard edge into eigenvalue
spectrum.

In the unitary case $\left( \beta =2\right) $, which applies to physical
systems with broken time-reversal symmetry, the structure of Eq. (\ref{i.01}%
) allows one to represent {\it exactly} all the global and local statistical
characteristics of the physical system, such as averaged density of levels, $%
n$-point correlation functions, level-spacing distribution function etc., in
terms of polynomials orthogonal with respect to the weight function $w\left(
x\right) =\exp \left\{ -2V\left( x\right) \right\} $ on the whole real axis 
{\bf R} (or on {\bf R}$^{+}$ if there is a hard edge in eigenvalue
spectrum). [Otherwise, when $\beta =1$ or $\beta =4$ more complicated sets
of skew orthogonal polynomials should be introduced \cite{Mahoux}].

Analytical calculation of the corresponding set of orthogonal polynomials is
a non-trivial problem. However, if the elements $H_{ij}$ of the random
matrix ${\bf H}$ are believed to be statistically independent from each
other, one obtains the quadratic confinement potential $V\left( x\right)
\sim x^2$ \cite{Porter} leading to the Gaussian Invariant Ensembles of
random matrices. In such a case there are significant mathematical
simplifications allowing to solve the matrix model Eq. (\ref{i.00})
completely \cite{Mehta}.

From the very beginning it was understood \cite{Dyson-1962} that requirement
of statistical independence of the matrix elements $H_{ij}$ is not motivated
by the first principles, and, therefore, several attempts were undertaken to
elucidate an influence of a particular form of confinement potential on the
predictions of the random matrix theory developed for Gaussian Ensembles.

Two essentially different lines of inquires of this problem can be
distinguished. The first line lies in the framework of polynomial approach,
while a second one consists of developing of a number of approximate
methods. The mean-field approximation proposed by Dyson \cite{Dyson} allows
to calculate density of levels in random-matrix ensemble. This approach
being combined with the method of functional derivative of Beenakker \cite
{Beenakker,Beenakker-2} makes it possible to compute global (smoothed)
eigenvalue correlations in large random matrices. Smoothed correlations can
also be obtained by diagrammatic approach of Br\'ezin and Zee \cite
{Brezin-Zee-diagrams} and by invoking the linear response arguments and
macroscopic electrostatics \cite{Forrester}. We stress that all the methods
mentioned above allow to study correlations only in {\it long-range regime}
and, in this sense, they are less informative as compared with the method of
orthogonal polynomials \cite{Mehta}. It is worth pointing out the
supersymmetry formalism \cite{Weidenmuller}, recently developed for matrix
model Eq. (\ref{i.00}) with non-Gaussian probability distribution function $%
P\left[ {\bf H}\right] $, which is exceptional in that it allows to
investigate {\it local} eigenvalue correlations and represents a powerful
alternative approach to the classical method of orthogonal polynomials.

In the framework of polynomial approach there was a number of studies to go
beyond the Gaussian distribution $P\left[ {\bf H}\right] $. In Refs. \cite
{Fox} - \cite{Bronk} it was found out that unitary random-matrix ensembles
associated with classical orthogonal polynomials exhibit Wigner-Dyson level
statistics [for corresponding ensembles with orthogonal and symplectic
symmetry see Ref. \cite{Nagao}]. Non-Gaussian unitary random-matrix
ensembles associated with (symmetric) strong confinement potentials $V\left(
x\right) =x^2+\gamma x^4$ and $V\left( x\right) =\sum_{n=1}^{n=p}a_nx^{2n}$
were treated in Refs. \cite{Mahoux} and \cite{Brezin-Zee}, respectively. (We
note that both potentials mentioned above are stronger than quadratic, and
they do not refer to the maximum entropy models). As far as these works have
been based on different {\it conjectures} about functional form of
asymptotics of polynomials orthogonal with respect to a non-Gaussian
measure, and the problem of hard edge in eigenvalue spectrum was out of
their scope, the polynomial approach to basic problems of the random matrix
theory needs further and more rigorous study.

The purpose of the present work is to show that the problem of non-Gaussian
ensembles with unitary symmetry does can be handled rigorously by the method
of orthogonal polynomials. Our treatment is exact (i.e. it does not involve
any conjectures) and based on the recent results obtained in the theory of
polynomials orthogonal with respect to exponential weights on {\bf R}. It
applies to very large class of confinement potentials which is much richer
than that considered in Refs. \cite{Mahoux,Brezin-Zee} and allows also to
treat the matrix models with positive constraints on eigenvalue spectrum. We
concentrate on the calculations of density of levels, one- and two-point
Green's functions, two-point kernel and connected ``density-density''
correlation function over {\it all distance scale}. This allows us to
resolve the problem of universality for local and global correlations of
random-matrix eigenvalues and to establish a new universal local
relationship for properly normalized and rescaled connected two-point
Green's function. One of the interesting points we would like to stress is
that the mean-field approximation, widely used in the theory of random
matrices, naturally appears in our treatment without any physical
speculations and turns out to be closely allied with structure of Szeg\"o
function entering strong pointwise asymptotics of orthogonal polynomials.

The paper is organized as follows. Section II\ contains a short introduction
to the theory of polynomials orthogonal with respect to the Freud weights.
The asymptotic formula for orthonormal ``wave function'' that we need in
later sections is given there. In Section III we calculate two-point kernel
and resolve the problem of universality of level statistics. The density of
levels and one-point Green's function are computed in Section IV. Connection
between structure of Szeg\"o function and mean-field equation is established
there as well. Section V is devoted to the calculation of the two-point
connected Green's function; corresponding new universal local expression is
given. Section VI contains generalizations of the results obtained in the
preceedings Sections for a wider class of random matrices characterized by
Erd\"os-type confinement potential. In Section VII we present a treatment of
the maximum entropy models with hard edge. Finally, in Section VIII we
discuss the results obtained.

\section{Freud-type confinement potentials and corresponding orthogonal
polynomials}

Let us consider a class of symmetric (even) confinement potentials $V\left(
x\right) $ supported on the whole real axis $x\in \left( -\infty ,+\infty
\right) $ which are {\it of smooth polynomial growth at infinity and
increase there at least as }$\left| x\right| ^{1+\delta }$ ($\delta $ is
arbitrary small positive number). More precisely, we demand $V\left(
x\right) $ and $d^2V/dx^2$ be continuous in $x\in \left( 0,+\infty \right) $%
, and $dV/dx>0$ in the same domain of variable $x$. We also assume that for
some $A>1$ and $B>1$ the inequality 
\begin{equation}
A\leq 1+x\frac{d^2V/dx^2}{dV/dx}\leq B  \label{eq.01}
\end{equation}
holds for $x\in \left( 0,+\infty \right) $, and also for $x$ positive and
large enough 
\begin{equation}
x^2\frac{\left| d^3V/dx^3\right| }{dV/dx}\leq const\text{.}  \label{eq.02}
\end{equation}
The class of potentials $V\left( x\right) $ satisfying all the above
requirements is said to be of the {\it Freud type} \cite{AAM-1993}. The
typical examples of the Freud potentials are (i) $V\left( x\right) =\left|
x\right| ^\alpha $ with $\alpha >1$, and (ii) $V\left( x\right) =\left|
x\right| ^\alpha \ln ^\beta \left( \gamma +x^2\right) $ with $\alpha >1$, $%
\beta \in {\bf R}$, and $\gamma $ large enough.

Now it is convenient to introduce a set of polynomials $P_n\left( x\right) $
orthogonal with respect to the Freud (non-Gaussian) measure $d\alpha _{{\cal %
F}}\left( x\right) =$ $w_{{\cal F}}\left( x\right) dx=\exp \left( -2V\left(
x\right) \right) dx$, 
\begin{equation}
\int_{-\infty }^{+\infty }P_n\left( x\right) P_m\left( x\right) d\alpha _{%
{\cal F}}\left( x\right) =\delta _{nm}\text{,}  \label{eq.03}
\end{equation}
for which the following basic result was obtained by D. S. Lubinsky \cite
{AAM-1993}: 
\[
\lim_{n\rightarrow \infty }\int_{-1}^{+1}d\lambda \left\{ \sqrt{a_n}%
P_n\left( a_n\lambda \right) -\left( \frac 2\pi \right) ^{1/2}\right. 
\]
\begin{equation}
\left. \times Re\left[ z^nD^{-2}\left( F_n;\frac 1z\right) \right] \right\}
^2w_{{\cal F}}\left( a_n\lambda \right) =0\text{.}  \label{eq.04}
\end{equation}
Here parametrization $z=e^{i\theta }$ and $\lambda =\cos \theta $ was used.

Szeg\"o function $D\left( g;z\right) $, appeared in Eq. (\ref{eq.04}), is of
fundamental importance in the whole theory of orthogonal polynomials \cite
{Szego-1921}, and takes the form 
\begin{equation}
D\left( g;z\right) =\exp \left( \frac 1{4\pi }\int_{-\pi }^{+\pi }d\varphi 
\frac{1+ze^{-i\varphi }}{1-ze^{-i\varphi }}\ln g\left( \varphi \right)
\right) \text{.}  \label{eq.06}
\end{equation}
The first argument of Szeg\"o function in Eq. (\ref{eq.04}) is 
\begin{equation}
F_n\left( \varphi \right) =\exp \left( -V\left( a_n\cos \varphi \right)
\right) \left| \sin \varphi \right| ^{1/2}\text{,}  \label{eq.07}
\end{equation}
and $a_n$ is the $n$-th Mhaskar-Rahmanov-Saff number being the positive root
of the integral equation \cite{MRS-numb} 
\begin{equation}
n=\frac{2a_n}\pi \int_0^1\frac{\lambda d\lambda }{\sqrt{1-\lambda ^2}}\left( 
\frac{dV}{dx}\right) _{x=a_n\lambda }\text{.}  \label{eq.05}
\end{equation}
(In what follows it will be seen that $a_n$ is none other than band edge for
eigenvalues of corresponding random-matrix ensemble.)

Equation (\ref{eq.04}) may be rewritten in the different form passing on to
the integration over $x=a_n\lambda $ (so that parametrization $x=a_n\cos
\theta $ takes place): 
\[
\lim_{n\rightarrow \infty }\int_{-a_n}^{+a_n}dx\left\{ P_n\left( x\right)
-\left( \frac 2{\pi a_n}\right) ^{1/2}\right. 
\]
\begin{equation}
\left. \times Re\left[ z^nD^{-2}\left( F_n;\frac 1z\right) \right] \right\}
^2w_{{\cal F}}\left( x\right) =0.  \label{eq.08}
\end{equation}
Analogously, Eq. (\ref{eq.05}) reads 
\begin{equation}
n=\frac 2\pi \int_0^{a_n}\frac{xdx}{\sqrt{a_n^2-x^2}}\frac{dV}{dx}\text{.}
\label{eq.09}
\end{equation}
Since from Eq. (\ref{eq.09}) it follows that $\lim_{n\rightarrow \infty
}a_n\neq 0$, we immediately conclude that expression in parentheses of Eq. (%
\ref{eq.08}) asymptotically tends to zero as $n\rightarrow \infty $ on the
interval of integration $\left| x\right| <a_n$. If one is not interested in
remainder term, we arrive at the asymptotic formula for orthogonal
polynomials of the Freud type: 
\begin{equation}
P_n\left( x\right) =\sqrt{\frac 2{\pi a_n}}Re\left[ z^nD^{-2}\left( F_n;%
\frac 1z\right) \right] ,\text{ }x\in \left( -a_n,+a_n\right) \text{.}
\label{eq.10}
\end{equation}

Szeg\"o function $D\left( g;e^{i\theta }\right) $ may be represented as \cite
{Szego-monograph} 
\begin{equation}
D\left( g;e^{i\theta }\right) =\sqrt{g\left( \theta \right) }\exp \left[
i\Gamma \left( g;\theta \right) \right] \text{,}  \label{eq.11}
\end{equation}
where 
\begin{equation}
\Gamma \left( g;\theta \right) =\frac 1{4\pi }\int_{-\pi }^{+\pi }d\varphi
\cot \left( \frac{\theta -\varphi }2\right) \left[ \ln g\left( \varphi
\right) -\ln g\left( \theta \right) \right] \text{.}  \label{eq.12}
\end{equation}
Making use of the representation of Eqs. (\ref{eq.11}) and (\ref{eq.12}),
noting that $F_n\left( -\varphi \right) =F_n\left( \varphi \right) $ and $%
\Gamma \left( F_n;-\theta \right) =-\Gamma \left( F_n;\theta \right) $, we
obtain 
\[
D\left( F_n;\frac 1z\right) =\exp \left( -\frac 12V\left( a_n\cos \theta
\right) \right) 
\]
\begin{equation}
\times \left| \sin \theta \right| ^{1/4}\exp \left[ -i\Gamma \left(
F_n;\theta \right) \right] \text{.}  \label{eq.13}
\end{equation}
Then, Eqs. (\ref{eq.10}) and (\ref{eq.13}) yield 
\[
P_n\left( a_n\cos \theta \right) =\sqrt{\frac 2{\pi a_n}} 
\]
\begin{equation}
\times \frac{\exp \left( V\left( a_n\cos \theta \right) \right) }{\left|
\sin \theta \right| ^{1/2}}\cos \left( n\theta +\Gamma \left( F_n^2;\theta
\right) \right) \text{,}  \label{eq.14}
\end{equation}
where\widetext
\[
\Gamma \left( F_n^2;\theta \right) =\frac 1{4\pi }\int_{-\pi }^{+\pi
}d\varphi \cot \left( \frac{\theta -\varphi }2\right) \left[ \ln F_n^2\left(
\varphi \right) -\ln F_n^2\left( \theta \right) \right] 
\]
\[
=\frac 1{4\pi }\int_0^\pi d\varphi \left[ \ln F_n^2\left( \varphi \right)
-\ln F_n^2\left( \theta \right) \right] \left\{ \cot \left( \frac{\theta
-\varphi }2\right) +\cot \left( \frac{\theta +\varphi }2\right) \right\} 
\label{eq.16} 
\]
\begin{equation}
=\frac 1{2\pi }\int_0^\pi d\varphi \left[ \ln F_n^2\left( \varphi \right)
-\ln F_n^2\left( \theta \right) \right] \frac{\sin \theta }{\cos \varphi
-\cos \theta }\text{.}  \label{eq.17}
\end{equation}
\narrowtext
\noindent Introducing the new variable of integration $\xi =a_n\cos \varphi $
and using parametrization $x=a_n\cos \theta $ ($\left| x\right| <a_n$), we
get 
\[
\gamma _n\left( x\right) =\left. \Gamma \left( F_n^2;\theta \right) \right|
_{x=a_n\cos \theta } 
\]
\begin{equation}
=\frac 1{2\pi }\int_{-a_n}^{+a_n}d\xi \frac{\sqrt{a_n^2-x^2}}{\sqrt{%
a_n^2-\xi ^2}}\frac{h\left( \xi \right) -h\left( x\right) }{\xi -x}
\label{eq.18}
\end{equation}
with 
\begin{equation}
h\left( \xi \right) =-2V\left( \xi \right) +\frac 12\ln \left[ 1-\left( 
\frac \xi {a_n}\right) ^2\right] \text{.}  \label{eq.19}
\end{equation}
Since for $\left| x\right| <a_n$%
\begin{equation}
{\cal P}\int_{-a_n}^{+a_n}\frac{d\xi }{\left( \xi -x\right) \sqrt{a_n^2-\xi
^2}}=0  \label{eq.20}
\end{equation}
(here ${\cal P}$ stands for principal value of an integral), Eq. (\ref{eq.18}%
) can be rewritten in the form 
\begin{equation}
\gamma _n\left( x\right) =\frac 1{2\pi }{\cal P}\int_{-a_n}^{+a_n}d\xi \frac{%
\sqrt{a_n^2-x^2}}{\sqrt{a_n^2-\xi ^2}}\frac{h\left( \xi \right) }{\xi -x}%
\text{.}  \label{eq.21}
\end{equation}
Then we obtain following asymptotic formula for orthonormal ``wave
functions'' $\psi _n\left( x\right) =P_n\left( x\right) \exp (-V\left(
x\right) )$ that we need in what follows: 
\[
\psi _n\left( x\right) =\sqrt{\frac 2{\pi a_n}}\left[ 1-\left( \frac x{a_n}%
\right) ^2\right] ^{-1/4} 
\]
\begin{equation}
\times \cos \left[ n\arccos \left( \frac x{a_n}\right) +\gamma _n\left(
x\right) \right] \text{.}  \label{eq.22}
\end{equation}
We remind that Eq. (\ref{eq.22}) is valid for $\left| x\right| <a_n$ in the
limit $n\rightarrow \infty $.

\section{Two-point kernel and universal eigenvalue correlations}

Two-point kernel allowing to calculate all the global and local
characteristics for the random-matrix ensembles is determined as \cite{Mehta}
\begin{equation}
K_n\left( x,y\right) =\frac{k_{n-1}}{k_n}\frac{\psi _n\left( y\right) \psi
_{n-1}\left( x\right) -\psi _n\left( x\right) \psi _{n-1}\left( y\right) }{%
y-x}\text{,}  \label{eq.23}
\end{equation}
where $k_n$ is a leading coefficient of the orthogonal polynomial $P_n\left(
x\right) $. Substitution of Eq. (\ref{eq.22}) into Eq. (\ref{eq.23}) yields
in the large-$n$ limit 
\[
K_n\left( x,y\right) =\frac 2{\pi a_n}\frac{k_{n-1}}{k_n} 
\]
\[
\times \frac 1{y-x}\left\{ \left[ 1-\left( \frac x{a_n}\right) ^2\right]
\left[ 1-\left( \frac y{a_n}\right) ^2\right] \right\} ^{-1/4} 
\]
\begin{equation}
\times \left( \cos \Phi _{n-1}\left( x\right) \cos \Phi _n\left( y\right)
-\cos \Phi _{n-1}\left( y\right) \cos \Phi _n\left( x\right) \right) \text{,}
\label{eq.25}
\end{equation}
where 
\begin{equation}
\Phi _n\left( x\right) =\gamma _n\left( x\right) +n\arccos \left( \frac x{a_n%
}\right) \text{.}  \label{eq.26}
\end{equation}
In Eq. (\ref{eq.25}) the fact was used that $\lim_{n\rightarrow \infty
}\left( a_{n-1}/a_n\right) =1$. Really, as was noted in previous section,
the Freud-type potentials exhibit a polynomial growth at infinity. Supposing
that at large positive $x$ potential $V\left( x\right) $ roughly behaves as $%
x^\rho $ $\left( \rho >1\right) $ we immediately obtain the estimate (see
Eq. (\ref{eq.09})) $a_n\rightarrow n^{1/\rho }$ as $n\rightarrow \infty $.
Then, obviously, $\lim_{n\rightarrow \infty }\left( a_{n-1}/a_n\right) =1$.
Taking into account this limit and carrying out the changing of integration
variable $\xi ^{\prime }=\xi a_n/a_{n-1}$ in Eq. (\ref{eq.21}) we easily
obtain that in the large-$n$ limit $\gamma _{n-1}\left( x\right) =\gamma
_n\left( x\right) $, and as a consequence 
\begin{equation}
\Phi _{n-1}\left( x\right) =\Phi _n\left( x\right) -\arccos \left( \frac x{%
a_n}\right) \text{.}  \label{eq.27}
\end{equation}
Now Eqs. (\ref{eq.25}) and (\ref{eq.27}) give us\widetext
\[
K_n\left( x,y\right) =\frac 1{\pi \left( y-x\right) }\left\{ \left[ 1-\left( 
\frac x{a_n}\right) ^2\right] \left[ 1-\left( \frac y{a_n}\right) ^2\right]
\right\} ^{-1/4} 
\]
\[
\times \left\{ \cos \Phi _n\left( x\right) \cos \Phi _n\left( y\right) \frac{%
x-y}{a_n}-\sin \Phi _n\left( y\right) \cos \Phi _n\left( x\right) \sqrt{%
1-\left( \frac y{a_n}\right) ^2}\right. 
\]
\begin{equation}
\left. +\sin \Phi _n\left( x\right) \cos \Phi _n\left( y\right) \sqrt{%
1-\left( \frac x{a_n}\right) ^2}\right\} \text{.}  \label{eq.28}
\end{equation}
\narrowtext
\noindent When deriving we have used identity $\lim_{n\rightarrow \infty
}k_{n-1}/k_na_n=1/2$ proved in Ref. \cite{LMS-1988}. We stress that Eq. (\ref
{eq.28}) is valid for arbitrary $x$ and $y$ lying within the band $\left(
-a_n,+a_n\right) $.

Equation (\ref{eq.28}) allows to determine smoothed (over the rapid
oscillations) connected correlations $\nu _c\left( x,y\right) $ of the
density of eigenvalues $\nu _n\left( x\right) $ \cite{Beenakker-2,Brezin-Zee}%
, 
\[
\nu _c\left( x,y\right) =\overline{\left\langle \nu _n\left( x\right) \nu
_n\left( y\right) \right\rangle } 
\]
\begin{equation}
-\overline{\left\langle \nu _n\left( x\right) \right\rangle \left\langle \nu
_n\left( y\right) \right\rangle }=-\overline{K_n^2\left( x,y\right) },\text{ 
}x\neq y  \label{eq.29}
\end{equation}
by averaging over intervals $\left| \Delta x\right| \ll a_n$ and $\left|
\Delta y\right| \ll a_n$ but still containing many eigenlevels. Direct
calculations yield simple universal relationship 
\[
\nu _c\left( x,y\right) = 
\]
\begin{equation}
-\frac 1{2\pi ^2}\frac{a_n^2-xy}{\left( x-y\right) ^2\sqrt{a_n^2-x^2}\sqrt{%
a_n^2-y^2}},\text{ }x\neq y  \label{eq.30}
\end{equation}
with dependence on the potential $V\left( x\right) $ only through the
endpoint $a_n$ of the spectrum.

Now we turn to the local properties of two-point kernel. Assuming that in
Eq. (\ref{eq.28}) $\left| x-y\right| \ll a_n$ and both $x$ and $y$ stay away
from the (soft) band edge $a_n$ we obtain 
\begin{equation}
K_n\left( x,y\right) =\frac{\sin \left( \Phi _n\left( x\right) -\Phi
_n\left( y\right) \right) }{\pi \left( y-x\right) }\text{,}  \label{eq.35}
\end{equation}
where $\Phi _n\left( x\right) $ is defined by Eq. (\ref{eq.26}). This
two-point kernel may be rewritten in locally universal form. Taking into
account the integral representation 
\begin{equation}
\Phi _n\left( x\right) =\frac 12\arccos \left( \frac x{a_n}\right) -\pi
\int_0^x\omega _{a_n}\left( \xi \right) d\xi +\frac \pi 4\left( 2n-1\right) 
\text{,}  \label{eq.333}
\end{equation}
\begin{equation}
\omega _{a_n}\left( x\right) =\frac 2{\pi ^2}{\cal P}\int_0^{a_n}\frac{\xi
d\xi }{\xi ^2-x^2}\frac{dV}{d\xi }\frac{\sqrt{a_n^2-x^2}}{\sqrt{a_n^2-\xi ^2}%
}\text{,}  \label{eq.444}
\end{equation}
proved in Appendix, we see that Eq. (\ref{eq.35}) may be rewritten as

\begin{equation}
K_n\left( x,y\right) =\frac{\sin \left( \pi \int_x^y\omega _{a_n}\left( \xi
\right) d\xi \right) }{\pi \left( y-x\right) }\text{.}  \label{eq.46}
\end{equation}
The characteristic scale of the changing of $\omega _{a_n}\left( \xi \right) 
$ is $\left( \omega _{a_n}^{-1}d\omega _{a_n}/d\xi \right) ^{-1}\sim a_n$,
so that for $\left| x-y\right| \ll a_n$ (that has been supposed in Eq. (\ref
{eq.35})) Eq. (\ref{eq.46}) is reduced to universal form 
\begin{equation}
K_n\left( x,y\right) =\frac{\sin \left[ \pi \overline{\nu }_n\left(
y-x\right) \right] }{\pi \left( y-x\right) }  \label{eq.47}
\end{equation}
with $\overline{\nu }_n=\omega _{a_n}\left( \frac{x+y}2\right) $ playing the
role of local density of levels. Correspondingly, local two-level cluster
function being rewritten in rescaled variables $s$ and $s^{\prime }$ 
\begin{equation}
Y_2\left( s,s^{\prime }\right) =\left( \frac{K_n^2\left( x,y\right) }{%
\left\langle \nu _n\left( x\right) \right\rangle \left\langle \nu _n\left(
y\right) \right\rangle }\right) _{%
{x=x\left( s\right)  \atopwithdelims.. y=y\left( s^{\prime }\right) }
}=\frac{\sin ^2\left[ \pi \left( s-s^{\prime }\right) \right] }{\left[ \pi
\left( s-s^{\prime }\right) \right] ^2}  \label{eq.48}
\end{equation}
proves universal Wigner-Dyson level statistics in the unitary random-matrix
ensemble with Freud-type confinement potentials (here $s=\overline{\nu }_nx$
and $s^{\prime }=\overline{\nu }_ny$ are the eigenvalues measured in the
local mean level-spacing).

\section{Density of levels and one-point Green's function}

Expression for density of levels defined as 
\begin{equation}
\left\langle \nu _n\left( x\right) \right\rangle =\left\langle \text{tr}%
\delta \left( x-{\bf H}\right) \right\rangle =K_n\left( x,x\right)
\label{eq.49}
\end{equation}
immediately follows from Eq. (\ref{eq.35}): 
\begin{equation}
\left\langle \nu _n\left( x\right) \right\rangle =-\frac 1\pi \frac{d\Phi _n%
}{dx}=\frac 1\pi \left( \frac n{\sqrt{a_n^2-x^2}}-\frac{d\gamma _n}{dx}%
\right) \text{,}  \label{eq.50}
\end{equation}
(see Eq. (\ref{eq.26})). Using Eqs. (\ref{eq.11}) and (\ref{eq.18}), and
parametrization $x=a_n\cos \theta $, we obtain the formula 
\[
\left\langle \nu _n\left( x=a_n\cos \theta \right) \right\rangle =\frac 1{%
\pi a_n\sin \theta } 
\]
\begin{equation}
\times \frac d{d\theta }\left[ \arg D\left( e^{-2V\left( a_n\cos \varphi
\right) }\left| \sin \varphi \right| ;e^{i\theta }\right) +n\theta \right] 
\text{,}  \label{eq.51}
\end{equation}
which establishes the connection between density of levels in random-matrix
ensemble with Freud-type confinement potential and Szeg\"o function for
corresponding set of orthogonal polynomials, Eq. (\ref{eq.06}).

Another representation of level density can be obtained from Eqs. (\ref
{eq.47}) and (\ref{eq.444}): 
\begin{equation}
\left\langle \nu _n\left( x\right) \right\rangle =\frac 2{\pi ^2}{\cal P}%
\int_0^{a_n}\frac{\xi d\xi }{\xi ^2-x^2}\frac{dV}{d\xi }\frac{\sqrt{a_n^2-x^2%
}}{\sqrt{a_n^2-\xi ^2}}\text{.}  \label{eq.52}
\end{equation}

This formula is rather interesting and deserves more attention. Considering
this expression as an equation for $dV/dx$ one can resolve it invoking the
theory of integral equations with Cauchy kernel \cite{Akhiezer}: 
\begin{equation}
{\cal P}\int_{-a_n}^{+a_n}\frac{\left\langle \nu _n\left( x^{\prime }\right)
\right\rangle }{x-x^{\prime }}dx^{\prime }=\frac{dV}{dx}\text{.}
\label{eq.52a}
\end{equation}
Thus, one can think that density of levels is a solution of integral
equation 
\begin{equation}
V\left( x\right) =\int_{-a_n}^{+a_n}dx^{\prime }\left\langle \nu _n\left(
x^{\prime }\right) \right\rangle \ln \left| x-x^{\prime }\right| +\mu
\label{eq.52b}
\end{equation}
with $\mu $ being ``chemical potential''. It is non more than famous
mean-field equation which, in our treatment, finally follows from the
asymptotic formula Eq. (\ref{eq.10}) for orthogonal polynomials. Quite
surprisingly, Szeg\"o function Eq. (\ref{eq.06}) turns out to be closely
related to the mean-field approximation by Dyson \cite{Dyson}.

Now we can easily calculate the one-point Green's function 
\[
G^p\left( x\right) =\left\langle \text{tr}\frac 1{x-{\bf H}+ip0}%
\right\rangle = 
\]
\begin{equation}
\int_{-a_n}^{+a_n}d\xi \frac 1{x-\xi +ip0}\left\langle \text{tr}\delta
\left( \xi -{\bf H}\right) \right\rangle \text{,}  \label{eq.53}
\end{equation}
where $p=\pm 1$. Last integral can be rewritten as 
\begin{equation}
G^p\left( x\right) ={\cal P}\int_{-a_n}^{+a_n}d\xi \frac{\left\langle \nu
_n\left( \xi \right) \right\rangle }{x-\xi }-i\pi p\left\langle \nu _n\left(
x\right) \right\rangle \text{,}  \label{eq.54}
\end{equation}
whence we obtain by means of Eqs. (\ref{eq.52}) and (\ref{eq.52a}): 
\begin{equation}
G^p\left( x\right) =\frac{dV}{dx}-\frac{2ip}\pi {\cal P}\int_0^{a_n}\frac{%
\xi d\xi }{\xi ^2-x^2}\frac{dV}{d\xi }\frac{\sqrt{a_n^2-x^2}}{\sqrt{%
a_n^2-\xi ^2}}\text{.}  \label{eq.55}
\end{equation}

We would like to stress that both Eqs. (\ref{eq.52}) and (\ref{eq.55}) have
been obtained within the framework of the theory of polynomials orthogonal
with respect to the Freud measure. This comment equally pertains to the
mean-field equation Eq. (\ref{eq.52b}).

\section{Two-point connected Green's function}

Two-point connected Green's function is defined as 
\[
G_c^{pp^{\prime }}\left( x,x^{\prime }\right) =\left\langle \text{tr}\frac 1{%
x_p-{\bf H}}\text{tr}\frac 1{x_{p^{\prime }}^{\prime }-{\bf H}}\right\rangle 
\]
\begin{equation}
-\left\langle \text{tr}\frac 1{x_p-{\bf H}}\right\rangle \left\langle \text{%
tr}\frac 1{x_{p^{\prime }}^{\prime }-{\bf H}}\right\rangle \text{,}
\label{eq.200}
\end{equation}
where $x_p=x+ip0$ and $x_{p^{\prime }}^{\prime }=x^{\prime }+ip^{\prime }0$ (%
$p,p^{\prime }=\pm 1$). It can be rewritten in an integral form 
\[
G_c^{pp^{\prime }}\left( x,x^{\prime }\right)
=\int_{-a_n}^{+a_n}\int_{-a_n}^{+a_n}\frac{d\xi d\eta }{\left( x_p-\xi
\right) \left( x_{p^{\prime }}^{\prime }-\eta \right) } 
\]
\begin{equation}
\times \left[ \left\langle \nu _n\left( \xi \right) \nu _n\left( \eta
\right) \right\rangle -\left\langle \nu _n\left( \xi \right) \right\rangle
\left\langle \nu _n\left( \eta \right) \right\rangle \right] \text{.}
\label{eq.201}
\end{equation}
Recognizing that the quantity in parentheses is $\left\langle \nu _n\left(
\xi \right) \nu _n\left( \eta \right) \right\rangle _c=\left\langle \nu
_n\left( \xi \right) \right\rangle \delta \left( \xi -\eta \right)
-K_n^2\left( \xi ,\eta \right) $, we obtain the formula 
\[
G_c^{pp^{\prime }}\left( x,x^{\prime }\right) = 
\]
\[
\int_{-a_n}^{+a_n}\frac{d\xi \left\langle \nu _n\left( \xi \right)
\right\rangle }{\left( x_p-\xi \right) \left( x_{p^{\prime }}^{\prime }-\xi
\right) }+\pi ^2pp^{\prime }K_n^2\left( x,x^{\prime }\right) 
\]
\begin{equation}
+i\pi \left[ p\Lambda \left( x,x^{\prime }\right) +p^{\prime }\Lambda \left(
x^{\prime },x\right) \right] -\lambda \left( x,x^{\prime }\right) \text{,}
\label{eq.202}
\end{equation}
where the following notations were used: 
\begin{equation}
\Lambda \left( x,x^{\prime }\right) ={\cal P}\int_{-a_n}^{+a_n}d\xi \frac{%
K_n^2\left( x,\xi \right) }{x^{\prime }-\xi }\text{,}  \label{eq.203}
\end{equation}
\begin{equation}
\lambda \left( x,x^{\prime }\right) ={\cal P}\int_{-a_n}^{+a_n}d\xi \frac{%
\Lambda \left( \xi ,x^{\prime }\right) }{x-\xi }\text{.}  \label{eq.204}
\end{equation}
Two-point kernel $K_n\left( x,x^{\prime }\right) $ entering Eqs. (\ref
{eq.202}), (\ref{eq.203}) is determined by Eq. (\ref{eq.28}).

\subsection{Smoothed connected two-point Green's function}

Let us consider the first integral Eq. (\ref{eq.203}). Substituting Eq. (\ref
{eq.28}) into Eq. (\ref{eq.203}) and taking into account that terms of the
type $\sin \Phi _n\left( \xi \right) $, $\cos \Phi _n\left( \xi \right) $,
and $\sin \Phi _n\left( \xi \right) \cos \Phi _n\left( \xi \right) $
oscillate rapidly and, therefore, do not contribute into integral over $\xi $
in the leading order in $n\gg 1$, we have after some rearrangements 
\[
\Lambda \left( x,x^{\prime }\right) =\frac 1{2\pi ^2}\frac 1{\sqrt{%
1-x^2/a_n^2}} 
\]
\begin{equation}
\times {\cal P}\int_{-a_n}^{+a_n}\frac{d\xi }{\left( x^{\prime }-\xi \right)
\left( x-\xi \right) ^2}\frac 1{\sqrt{1-\xi ^2/a_n^2}}\left( 1-\frac{x\xi }{%
a_n^2}\right)  \label{eq.206}
\end{equation}
provided $x\neq x^{\prime }$. Formally, this integral is divergent thanks to
the double pole of integrand $\propto \left( x-\xi \right) ^{-2}$. It is
easy to see that this singularity is rather artificial and connected with
the fact that condition $x\neq \xi $ was supposed to be fulfilled when
neglecting rapid oscillations in $\xi $ in integrand of Eq. (\ref{eq.203}).
This is the reason why the integrand in Eq. (\ref{eq.206}) displays a
non-correct behavior in the vicinity $x=\xi $. Actually, as can be verified,
the integrand is finite for $x=\xi $, and corresponding integral is
convergent. Moreover, direct comparison of Eq. (\ref{eq.206}) with results 
\cite{Beenakker-2} shows that equation in question can be rewritten in the
form 
\begin{equation}
\Lambda \left( x,x^{\prime }\right) ={\cal P}\int_{-a_n}^{+a_n}\frac{d\xi }{%
x^{\prime }-\xi }T_2\left( \xi ,x\right) \text{,}  \label{eq.207}
\end{equation}
where 
\begin{equation}
T_2\left( \xi ,x\right) =K_2\left( \xi ,x\right) +\left\langle \nu _n\left(
\xi \right) \right\rangle \delta \left( \xi -x\right)  \label{eq.208}
\end{equation}
is two-level cluster function, and 
\begin{equation}
K_2\left( \xi ,x\right) =\frac 12\frac{\delta \left\langle \nu _n\left( \xi
\right) \right\rangle }{\delta V\left( x\right) }  \label{eq.209}
\end{equation}
is two-point correlation function (the notations of Ref. \cite{Beenakker-2}
have been used). Then, taking into account Eqs. (\ref{eq.207}), (\ref{eq.208}%
) and (\ref{eq.204}), we obtain from Eq. (\ref{eq.202}) after some
transformations: 
\[
\overline{G_c^{pp^{\prime }}\left( x,x^{\prime }\right) }=\pi ^2pp^{\prime }%
\overline{K_n^2\left( x,x^{\prime }\right) }+i\pi 
\]
\[
\times \left[ p{\cal P}\int_{-a_n}^{+a_n}\frac{d\xi }{x^{\prime }-\xi }%
K_2\left( \xi ,x\right) +p^{\prime }{\cal P}\int_{-a_n}^{+a_n}\frac{d\xi }{%
x-\xi }K_2\left( \xi ,x^{\prime }\right) \right] \label{eq.202} 
\]
\begin{equation}
-{\cal PP}\int_{-a_n}^{+a_n}\int_{-a_n}^{+a_n}\frac{d\xi d\eta }{\left(
x-\xi \right) \left( x^{\prime }-\eta \right) }K_2\left( \xi ,\eta \right) 
\text{.}  \label{eq.210}
\end{equation}
Now we only have to calculate the integrals containing $K_2$. The most
proper way is to invoke the integral equation \cite{Beenakker-2} 
\begin{equation}
{\cal P}\int_{-a_n}^{+a_n}\frac{d\xi }{x-\xi }\delta \left\langle \nu
_n\left( \xi \right) \right\rangle =\frac d{dx}\delta V\left( x\right)
\label{eq.211}
\end{equation}
and definition Eq. (\ref{eq.209}). Since Eqs. (\ref{eq.209}), (\ref{eq.211})
yield identity 
\[
i\pi \left[ p{\cal P}\int_{-a_n}^{+a_n}\frac{d\xi }{x^{\prime }-\xi }%
K_2\left( \xi ,x\right) +p^{\prime }{\cal P}\int_{-a_n}^{+a_n}\frac{d\xi }{%
x-\xi }K_2\left( \xi ,x^{\prime }\right) \right] 
\]
\[
-{\cal PP}\int_{-a_n}^{+a_n}\int_{-a_n}^{+a_n}\frac{d\xi d\eta }{\left(
x-\xi \right) \left( x^{\prime }-\eta \right) }K_2\left( \xi ,\eta \right) 
\]
\begin{equation}
=-\frac 12\frac 1{\left( x_p-x_{p^{\prime }}^{\prime }\right) ^2}\text{,}
\label{eq.212}
\end{equation}
we finally arrive at the expression for two-point connected Green's
function: 
\[
\overline{G_c^{pp^{\prime }}\left( x,x^{\prime }\right) }=\frac 12 
\]
\begin{equation}
\times \left\{ pp^{\prime }\frac{a_n^2-xx^{\prime }}{\left( x-x^{\prime
}\right) ^2\sqrt{a_n^2-x^2}\sqrt{a_n^2-x^{\prime 2}}}-\frac 1{\left(
x_p-x_{p^{\prime }}^{\prime }\right) ^2}\right\} \text{.}  \label{eq.215}
\end{equation}
Here we have used Eqs. (\ref{eq.29}) and (\ref{eq.30}). Equation (\ref
{eq.215}) is valid for arbitrary $x\neq x^{\prime }$ lying within the band $%
\left( -a_n,+a_n\right) $. Universal relationships of this type were
obtained in Ref. \cite{Brezin-Zee}.

\subsection{Local connected two-point Green's function}

In the local regime, when $\left| x-x^{\prime }\right| \ll a_n$, one cannot
disregard oscillations of integrands in Eqs. (\ref{eq.203}) and (\ref{eq.204}%
). Since in this energy scale the density of states $\left\langle \nu
_n\left( x\right) \right\rangle $ is slowly varying function and two-point
kernel $K_n\left( x,x^{\prime }\right) $ is universal, Eq. (\ref{eq.47}),
one obtains that \cite{Bateman} 
\begin{equation}
\Lambda \left( x,x^{\prime }\right) =\frac{\overline{\nu }_n}{x^{\prime }-x}%
\left\{ 1-\frac{\sin \left[ 2\pi \overline{\nu }_n\left( x^{\prime
}-x\right) \right] }{2\pi \overline{\nu }_n\left( x^{\prime }-x\right) }%
\right\} \text{,}  \label{eq.777}
\end{equation}
and 
\begin{equation}
\lambda \left( x,x^{\prime }\right) =\frac{\sin ^2\left[ \pi \overline{\nu }%
_n\left( x-x^{\prime }\right) \right] }{\left( x-x^{\prime }\right) ^2}.
\label{eq.778}
\end{equation}
Then Eqs. (\ref{eq.777}), (\ref{eq.778}) and (\ref{eq.202}) yield 
\[
G_c^{pp^{\prime }}\left( x,x^{\prime }\right) =\pi ^2\overline{\nu }_n\left|
p-p^{\prime }\right| \delta \left( x-x^{\prime }\right) 
\]
\[
+\left[ pp^{\prime }-1\right] \frac{\sin ^2\left[ \pi \overline{\nu }%
_n\left( x-x^{\prime }\right) \right] }{\left( x-x^{\prime }\right) ^2} 
\]
\begin{equation}
+i\left( p^{\prime }-p\right) \frac{\sin \left[ \pi \overline{\nu }_n\left(
x^{\prime }-x\right) \right] \cos \left[ \pi \overline{\nu }_n\left(
x^{\prime }-x\right) \right] }{\left( x^{\prime }-x\right) ^2}.
\label{eq.779}
\end{equation}
This equation only depends on the local mean-level spacing $\overline{\nu }%
_n,$ and therefore it can be written down in universal form. Introducing
normalized and rescaled two-point connected Green's function 
\begin{equation}
g_c^{pp^{\prime }}\left( s,s^{\prime }\right) =\left( \frac{G_c^{pp^{\prime
}}\left( x,x^{\prime }\right) }{\left\langle \nu _n\left( x\right)
\right\rangle \left\langle \nu _n\left( x^{\prime }\right) \right\rangle }%
\right) _{%
{x=x\left( s\right)  \atopwithdelims.. x^{\prime }=x^{\prime }\left( s^{\prime }\right) }
}\text{,}  \label{eq.216}
\end{equation}
where $s=\overline{\nu }_nx$ and $s^{\prime }=\overline{\nu }_nx^{\prime }$
are the eigenvalues measured in the local mean level-spacing, we obtain new
universal relationship in the random-matrix theory: 
\[
g_c^{pp^{\prime }}\left( s,s^{\prime }\right) =\pi ^2\left| p-p^{\prime
}\right| \delta \left( s-s^{\prime }\right) 
\]
\begin{equation}
+i\left( p-p^{\prime }\right) \frac{\sin \left[ \pi \left( s-s^{\prime
}\right) \right] }{\left( s-s^{\prime }\right) ^2}\text{e}^{i\pi \left(
s-s^{\prime }\right) \text{sign}\left( p-p^{\prime }\right) }\text{.}
\label{eq.217}
\end{equation}
Note that expression of this type was previously obtained in Ref. \cite{Zuk}
only for Gaussian random-matrix ensemble using supersymmetry formalism.

\section{Extension for Erd\"os-type confinement potentials}

All the results obtained above are valid for confinement potentials
exhibiting smooth polynomial growth at infinity (see Section II) but they
can be extended for an {\it Erd\"os-type }confinement potentials which {\it %
grow faster than any polynomial at infinity} (see Ref. \cite{PRN}, Ch. 2).

Namely, let $V\left( x\right) $ be even and continuous in $x\in \left(
-\infty ,+\infty \right) $, $d^2V/dx^2$ be continuous in $x\in \left(
0,+\infty \right) $, $dV/dx$ be positive in the same domain of $x$ and
continuous at $x=0$. Moreover, let 
\begin{equation}
T\left( x\right) =1+x\frac{d^2V/dx^2}{dV/dx}  \label{eq.60}
\end{equation}
be positive and increasing in $x\in \left( 0,+\infty \right) $ with $%
\lim_{x\rightarrow +0}T\left( x\right) >0$ while $\lim_{x\rightarrow \infty
}T\left( x\right) =\infty $, and 
\begin{equation}
T\left( x\right) ={\cal O}\left( \left( dV/dx\right) ^{1/15}\right) \text{
for }x\rightarrow \infty \text{,}  \label{eq.61}
\end{equation}
\begin{equation}
\frac{d^2V/dx^2}{dV/dx}\sim \frac{dV/dx}{V\left( x\right) }\text{ and }\frac{%
\left| d^3V/dx^3\right| }{dV/dx}\leq const\cdot \left( \frac{dV/dx}{V\left(
x\right) }\right) ^2  \label{eq.62}
\end{equation}
for $x$ large enough. The class of potentials $V\left( x\right) $ satisfying
all the above requirements is said to be of the {\it Erd\"os-type}. The
simple examples of Erd\"os-type confinement potentials are (i) $V\left(
x\right) =\exp _k\left( \left| x\right| ^\alpha \right) $ with $\alpha >0$
and $k\geq 1$ (here $\exp _k$ denotes the exponent iterated $k$-times); (ii) 
$V\left( x\right) =\exp \left( \ln ^\alpha \left( \gamma +x^2\right) \right) 
$ with $\alpha >1$, and $\gamma $ large enough.

Polynomials orthogonal with respect to the Erd\"os measure $d\alpha _{{\cal E%
}}=w_{{\cal E}}\left( x\right) dx=\exp \left( -2V\left( x\right) \right) dx$
(here $V$ is of Erd\"os type) have the same asymptotics \cite{PRN} and,
therefore, Eq. (\ref{eq.22}) remains valid along with all the results
obtained in Sections III, IV and V.

\section{Matrix models with positivity constraints on eigenvalues}

In the random-matrix theory of quantum transport \cite
{Mesoscopics,Slevin-Nagao} the matrix model Eq. (\ref{i.01}) appears with
positivity constraints on eigenvalues $\left\{ x\right\} $ (maximum entropy
models). The constraint $x\geq 0$ is essential feature of those models that
follows directly from the unitarity of scattering matrix and imposes the
presence of the hard edge in the energy spectrum of the matrix model. To our
knowledge there is no rigorous treatment of such matrix model with strong
enough confinement potential $V\left( x\right) $ within the method of
orthogonal polynomials except for generalized Laguerre ensembles of random
matrices \cite{Laguerre}.

Below we show how the problems associated with maximum entropy model can be
treated within the polynomial approach in very general case.

\subsection{Polynomials orthogonal on $x\geq 0$}

Let confinement potential $V\left( x\right) $ be of the Freud- or
Erd\"os-type defined on the whole real axis {\bf R}, that is $V$ is
monotonous function behaving at least as $\left| x\right| ^{1+\delta }$ $%
\left( \delta >0\right) $ and growing as or even faster than any polynomial
at infinity, and $P_n\left( x\right) $ be a set of polynomials orthogonal on 
{\bf R} with respect to the measure $d\alpha \left( x\right) =$ $\exp
\left\{ -2V\left( x\right) \right\} dx$ (see Eq. (\ref{eq.03})). Then
polynomials 
\begin{equation}
S_n\left( x\right) =P_{2n}\left( \sqrt{x}\right)  \label{eq.887}
\end{equation}
form a set of polynomials orthogonal on {\bf R}$^{{\bf +}}$ with the measure 
\cite{Chihara} $d\alpha _s\left( x\right) =\exp \left\{ -2V_s\left( x\right)
\right\} dx$, 
\begin{equation}
\int_0^\infty S_n\left( x\right) S_m\left( x\right) d\alpha _s\left(
x\right) =\delta _{nm}\text{,}  \label{eq.888}
\end{equation}
where confinement potential 
\begin{equation}
V_s\left( x\right) =V\left( \sqrt{x}\right) +\frac 14\ln x  \label{eq.889}
\end{equation}
is a monotonous function that behaves at least as $\left| x\right| ^{\frac 12%
+\delta }$ $\left( \delta >0\right) $ and can grow even faster than any
polynomial at infinity.

Equation (\ref{eq.887}) allows to write down large-$n$ asymptotics for
introduced set of orthogonal polynomials. It is straightforward to get from
the results outlined in Section II and Appendix the following asymptotic
formula [which is analogue of Eq. (\ref{eq.14})]: 
\begin{equation}
S_n\left( x\right) =\sqrt{\frac 2\pi }\frac{\exp \left( V_s\left( x\right)
\right) }{\left( xb_n\right) ^{1/4}}\frac 1{\left[ 1-x/b_n\right] ^{1/4}}%
\cos \widetilde{\Phi }_n\left( x\right) ,\text{ }  \label{eq.890}
\end{equation}
where $x\in \left( 0,b_n\right) ,$ and 
\[
\widetilde{\Phi }_n\left( x\right) =\frac 12\arccos \left( \sqrt{\frac x{b_n}%
}\right) 
\]
\begin{equation}
+\pi \left( n-\frac 14\right) -\pi \int_0^x\Omega _{b_n}\left( \xi \right)
d\xi \text{,}  \label{eq.891}
\end{equation}
\begin{equation}
\Omega _{b_n}\left( x\right) =\frac 1{\pi ^2}{\cal P}\int_0^{b_n}\frac{d\eta 
}{\eta -x}\frac{dV_s}{d\eta }\sqrt{\frac \eta x}\frac{\sqrt{b_n-x}}{\sqrt{%
b_n-\eta }}\text{.}  \label{eq.892}
\end{equation}
Here soft band edge $b_n=a_{2n}^2$.

Equations obtained above are the starting point of the further analysis.

\subsection{Two-point kernel and universal eigenvalue correlations}

Two-point kernel determined by Eq. (\ref{eq.23}) can be calculated provided
``wave function'' $\psi _n\left( x\right) =\exp \left( -V_s\left( x\right)
\right) S_n\left( x\right) $. Substitution of Eq. (\ref{eq.890}) into Eq. (%
\ref{eq.23}) yields in the large-$n$ limit 
\[
K_n\left( x,y\right) =\frac 2\pi \frac{\widetilde{k}_{n-1}}{\widetilde{k}_n} 
\]
\[
\times \frac 1{\left( y-x\right) }\frac 1{\left( xy\right) ^{1/4}\left\{
\left[ b_n-x\right] \left[ b_n-y\right] \right\} ^{1/4}} 
\]
\begin{equation}
\times \left( \cos \widetilde{\Phi }_{n-1}\left( x\right) \cos \widetilde{%
\Phi }_n\left( y\right) -\cos \widetilde{\Phi }_{n-1}\left( y\right) \cos 
\widetilde{\Phi }_n\left( x\right) \right)  \label{eq.893}
\end{equation}
if $x$ and $y$ lie within the band $\left( 0,b_n\right) $. If at least one
of the arguments in two-point kernel is negative, it is identically zero
(due to presence of hard edge). In Eq. (\ref{eq.893}) $\widetilde{k}_n$
stands for leading coefficient of $S_n\left( x\right) $.

Taking into account the large-$n$ identity 
\begin{equation}
\widetilde{\Phi }_{n-1}\left( x\right) =\widetilde{\Phi }_n\left( x\right)
-2\arccos \left( \sqrt{\frac x{b_n}}\right) \text{,}  \label{eq.894}
\end{equation}
we obtain \widetext
\[
K_n\left( x,y\right) =\frac 4\pi \frac{\widetilde{k}_{n-1}}{\widetilde{k}_n}%
\frac 1{\left( y-x\right) }\frac 1{\left( xy\right) ^{1/4}\left\{ \left[
b_n-x\right] \left[ b_n-y\right] \right\} ^{1/4}} 
\]
\[
\times \left\{ \cos \widetilde{\Phi }_n\left( x\right) \cos \widetilde{\Phi }%
_n\left( y\right) \frac{x-y}{b_n}-\sin \widetilde{\Phi }_n\left( y\right)
\cos \widetilde{\Phi }_n\left( x\right) \sqrt{\frac y{b_n}}\sqrt{1-\frac y{%
b_n}}\right. 
\]
\begin{equation}
\left. +\sin \widetilde{\Phi }_n\left( x\right) \cos \widetilde{\Phi }%
_n\left( y\right) \sqrt{\frac x{b_n}}\sqrt{1-\frac x{b_n}}\right\} \text{.}
\label{eq.895}
\end{equation}

\narrowtext
\noindent Smoothed (over the rapid oscillations) connected correlator $\nu
_c\left( x,y\right) $ of the density of eigenvalues, Eq. (\ref{eq.29}), 
\[
\nu _c\left( x,y\right) = 
\]
\begin{equation}
-\frac 1{2\pi ^2}\frac{b_n\left( x+y\right) /2-xy}{\left( x-y\right) ^2\sqrt{%
xy}\sqrt{b_n-x}\sqrt{b_n-y}},\text{ }x\neq y  \label{eq.896}
\end{equation}
manifests dependence on the potential $V\left( x\right) $ only through the
soft edge $b_n$ of the spectrum.

The local properties of two-point kernel are obtained by assuming that in
Eq. (\ref{eq.895}) $\left| x-y\right| \ll b_n$ and both $x$ and $y$ stay
away from the hard edge $x=0$ and soft edge $x=b_n$:

\begin{equation}
K_n\left( x,y\right) =\frac{\sin \left( \pi \int_x^y\Omega _{b_n}\left( \xi
\right) d\xi \right) }{\pi \left( y-x\right) }\text{.}  \label{eq.897}
\end{equation}
The characteristic scale of the changing of $\Omega _{b_n}\left( \xi \right) 
$ is of the order of $b_n$, so that for $\left| x-y\right| \ll b_n$ Eq. (\ref
{eq.897}) is reduced to universal form Eq. (\ref{eq.47}) with $\overline{\nu 
}_n=\Omega _{b_n}\left( \frac{x+y}2\right) $ playing the role of local
density of levels. Correspondingly, local two-level cluster function $%
Y_2\left( s,s^{\prime }\right) $ being rewritten in rescaled variables $s$
and $s^{\prime }$ follows the universal form Eq. (\ref{eq.48}) that proves
universal Wigner-Dyson level statistics in the bulk of the spectrum for
unitary random-matrix ensembles with confinement potentials $V_s\left(
x\right) $.

\subsection{Density of levels and one-point Green's function}

Density of levels is obtained from Eq. (\ref{eq.897}) in the limit $%
y\rightarrow x$: 
\begin{equation}
\left\langle \nu _n\left( x\right) \right\rangle =\frac 1{\pi ^2}{\cal P}%
\int_0^{b_n}\frac{d\eta }{\eta -x}\frac{dV_s}{d\eta }\sqrt{\frac \eta x}%
\frac{\sqrt{b_n-x}}{\sqrt{b_n-\eta }}.  \label{eq.898}
\end{equation}
Considering this expression as an equation for $dV_s/dx$ one can resolve it 
\cite{Akhiezer} arriving to the mean-field equation by Dyson 
\begin{equation}
V_s\left( x\right) =\int_0^{b_n}dx^{\prime }\left\langle \nu _n\left(
x^{\prime }\right) \right\rangle \ln \left| x-x^{\prime }\right| +\mu \text{,%
}  \label{eq.899}
\end{equation}
where integration runs over $x^{\prime }\in \left( 0,b_n\right) $. We once
more stress that the mean-field equation is a direct consequence of the
point-wise asymptotics for corresponding orthogonal polynomials $S_n\left(
x\right) $ which involve Szeg\"o function as a starting point.

Correspondingly, the one-point Green's function 
\begin{equation}
G^p\left( x\right) =\frac{dV_s}{dx}-\frac{ip}\pi {\cal P}\int_0^{b_n}\frac{%
d\eta }{\eta -x}\frac{dV_s}{d\eta }\sqrt{\frac \eta x}\frac{\sqrt{b_n-x}}{%
\sqrt{b_n-\eta }}\text{.}  \label{eq.900}
\end{equation}

\subsection{Connected two-point Green's function}

In the maximum entropy models the smoothed connected two-point Green's
function can be calculated in the same way as it was done in Section V. The
only difference is that integrals in Eqs. (\ref{eq.202}) - (\ref{eq.204}), (%
\ref{eq.210}) now run from $0$ to $b_n$. Carrying out this integration with
two-point kernel $K_n\left( x,y\right) $ given by Eq. (\ref{eq.895}) we
arrive at the universal formula 
\[
\overline{G_c^{pp^{\prime }}\left( x,x^{\prime }\right) }=\frac 12\left\{
pp^{\prime }\frac{\frac{b_n}2\left( x+x^{\prime }\right) -xx^{\prime }}{%
\left( x-x^{\prime }\right) ^2\sqrt{xx^{\prime }}\sqrt{b_n-x}\sqrt{b_n-x}}%
\right. 
\]
\begin{equation}
\left. -\frac 1{\left( x_p-x_{p^{\prime }}^{\prime }\right) ^2}\right\} 
\text{.}  \label{eq.901}
\end{equation}

In contrast to the smoothed connected two-point Green's function the local
one is determined by the same formulae Eqs. (\ref{eq.779}) - (\ref{eq.217})
provided $x$ and $y$ are far from both edges.

\section{Conclusion}

We have presented rigorous analytical consideration of the matrix model
given by non-Gaussian distribution function $P\left( \left\{ x\right\}
\right) $, Eq. (\ref{i.01}), with very general class of confinement
potentials $V\left( x\right) $ within the framework of
orthogonal-polynomials technique. Our treatment is equally applied to the
random matrix models with presence and absence of the hard edge in the
eigenvalue spectrum. We have calculated with asymptotic accuracy the density
of levels, one-point Green's function, two-point kernel, ``density-density''
correlator and two-point Green's function over all distance scale.

It was established that two-point correlators in considered random-matrix
model posses a high degree of universality. In the absence of hard edge the
universality is observed for very wide class of monotonous confinement
potentials $V\left( x\right) $ which behave at least as $\left| x\right|
^{1+\delta }$ $\left( \delta >0\right) $ and {\it can grow as or even faster
than any polynomial at infinity} (the case of border level confinement when $%
V\left( x\right) \sim \left| x\right| $ as $\left| x\right| \rightarrow
\infty $ has been treated in Ref. \cite{FKY}). In the presence of hard edge
in eigenvalue spectrum the universality holds for monotonous confinement
potentials $V_s\left( x\right) $ which behave at least as $\left| x\right| ^{%
\frac 12+\delta }$ $\left( \delta >0\right) $ and {\it can grow faster than
any polynomial at infinity.}

We have shown that in those unitary non-Gaussian random-matrix models the
density of levels and one-point Green's function essentially depend on the
measure, i. e. on the explicit form of confinement potential. In contrast,
(connected) two-point characteristics of spectrum (``density-density''
correlator, two-point Green's function) are rather universal. Indeed, we
have observed global universality of smoothed two-point connected
correlators and local universality of those without smoothing over rapid
oscillations. In both cases the correlators were shown to depend on the
measure only through the endpoints of spectrum (global universality) or
through the local density of levels (local universality).

Rigorous polynomial analysis enabled us to recover the results obtained
before by different approximate methods and to extend previously known
results for much wider class of random-matrix ensembles with strong
confinement potentials irrespective of presence/absence of hard edge. We
also have established a new local universal relationship in the
random-matrix theory for normalized and rescaled connected two-point Green's
function $g_c^{pp^{\prime }}\left( s,s^{\prime }\right) $ [see Eq. (\ref
{eq.217})]. Finally, it is worthy of notice an interesting and quite
surprising intimate connection between the structure of Szeg\"o function and
mean-field equation that has been revealed in the proposed formalism.

\begin{center}
{\bf ACKNOWLEDGMENT}
\end{center}

One of the authors (E. K.) gratefully acknowledges financial support from
The Ministry of Science and The Arts of Israel.\widetext
\newpage\ 

\begin{center}
{\bf Appendix: Integral representation of }$\Phi _n\left( x\right) $
\end{center}

To prove Eq. (\ref{eq.333}) let us calculate the first derivative of $\gamma
_n\left( x\right) $ (the calculations are similar to those done in Ref. \cite
{PRN}, Ch. 11). From Eq. (\ref{eq.21}) we obtain 
\[
\frac{d\gamma _n}{dx}=-\frac 1{2\pi }\frac x{\sqrt{a_n^2-x^2}}{\cal P}%
\int_{-a_n}^{+a_n}\frac{h\left( \xi \right) d\xi }{\sqrt{a_n^2-\xi ^2}\left(
\xi -x\right) } 
\]
\begin{equation}
+\frac 1{2\pi }\sqrt{a_n^2-x^2}{\cal P}\int_{-a_n}^{+a_n}\frac{h\left( \xi
\right) d\xi }{\sqrt{a_n^2-\xi ^2}\left( \xi -x\right) ^2}\text{,} 
\eqnum{A1}
\end{equation}
whence 
\[
\sqrt{a_n^2-x^2}\frac{d\gamma _n}{dx}=\frac 1{2\pi }{\cal P}%
\int_{-a_n}^{+a_n}\frac{h\left( \xi \right) d\xi }{\xi -x}\left( \frac \xi {%
\sqrt{a_n^2-\xi ^2}}+\frac{\sqrt{a_n^2-\xi ^2}}{\xi -x}\right) 
\]
\begin{equation}
=-\frac 1{2\pi }{\cal P}\int_{-a_n}^{+a_n}h\left( \xi \right) d\xi \frac d{%
d\xi }\left( \frac{\sqrt{a_n^2-\xi ^2}}{\xi -x}\right) \text{.}  \eqnum{A2}
\end{equation}
After integration by parts we have 
\begin{equation}
\frac{d\gamma _n}{dx}=\frac 1{\pi \sqrt{a_n^2-x^2}}{\cal P}\int_0^{a_n}d\xi 
\frac{\sqrt{a_n^2-\xi ^2}}{\xi ^2-x^2}\xi \frac{dh}{d\xi }\text{.} 
\eqnum{A3}
\end{equation}
Substituting Eq. (\ref{eq.19}) into Eq. (A3) and using identity 
\begin{equation}
{\cal P}\int_0^{a_n}\frac{d\xi }{\xi ^2-x^2}\frac 1{\sqrt{a_n^2-\xi ^2}}=0 
\eqnum{A4}
\end{equation}
we obtain 
\begin{equation}
\frac{d\gamma _n}{dx}=-\frac 2{\pi \sqrt{a_n^2-x^2}}{\cal P}\int_0^{a_n}d\xi 
\frac{\sqrt{a_n^2-\xi ^2}}{\xi ^2-x^2}\xi \frac{dV}{d\xi }-\frac 1{2\sqrt{%
a_n^2-x^2}}\text{.}  \eqnum{A5}
\end{equation}
The integral in Eq. (A5) may be handled as follows 
\[
{\cal P}\int_0^{a_n}d\xi \frac{\sqrt{a_n^2-\xi ^2}}{\xi ^2-x^2}\xi \frac{dV}{%
d\xi }={\cal P}\int_0^{a_n}\frac{\xi d\xi }{\xi ^2-x^2}\frac{dV}{d\xi }\frac{%
\sqrt{a_n^2-x^2}}{\sqrt{a_n^2-\xi ^2}} 
\]
\begin{equation}
-\frac 1{\sqrt{a_n^2-x^2}}\int_0^{a_n}\frac{\xi d\xi }{\sqrt{a_n^2-\xi ^2}}%
\frac{dV}{d\xi }\text{.}  \eqnum{A6}
\end{equation}
Bearing in mind Eq. (\ref{eq.09}) and introducing function 
\begin{equation}
\omega _{a_n}\left( x\right) =\frac 2{\pi ^2}{\cal P}\int_0^{a_n}\frac{\xi
d\xi }{\xi ^2-x^2}\frac{dV}{d\xi }\frac{\sqrt{a_n^2-x^2}}{\sqrt{a_n^2-\xi ^2}%
}  \eqnum{A7}
\end{equation}
the derivative $d\gamma _n/dx$ can be rewritten as 
\begin{equation}
\frac{d\gamma _n}{dx}=-\pi \omega _{a_n}\left( x\right) +\left( n-\frac 12%
\right) \frac 1{\sqrt{a_n^2-x^2}}\text{.}  \eqnum{A8}
\end{equation}
Further, noting from Eq. (\ref{eq.21}) that $\gamma _n\left( 0\right) =0$,
we obtain the integral representation 
\begin{equation}
\gamma _n\left( x\right) =-\pi \int_0^x\omega _{a_n}\left( \xi \right) d\xi
+\left( n-\frac 12\right) \arcsin \left( \frac x{a_n}\right) \text{,} 
\eqnum{A9}
\end{equation}
or, equivalently (see Eq. (\ref{eq.26})), 
\begin{equation}
\Phi _n\left( x\right) =\frac 12\arccos \left( \frac x{a_n}\right) -\pi
\int_0^x\omega _{a_n}\left( \xi \right) d\xi +\frac \pi 4\left( 2n-1\right) 
\text{.}  \eqnum{A10}
\end{equation}

\newpage\ \narrowtext
\noindent 

$^{*}$ On leave from Institute for Low Temperature Physics and Engineering,
Kharkov 310164, Ukraine

\widetext

\end{document}